\documentclass[prb, preprint, superscriptaddress, endfloats*]{revtex4-1}
\usepackage{chemformula} 
\usepackage[T1]{fontenc} 
\usepackage{graphicx}
\usepackage{verbatim}
\usepackage{mathrsfs}
\usepackage{gensymb}
\usepackage{amsmath,amssymb,amsfonts,bm}

\begin{document}

\title{Observation of Unpinned Two-Dimensional Dirac States in Antimony Single Layers with Phosphorene Structure}

\author{Qiangsheng Lu}
\affiliation {Department of Physics and Astronomy, University of Missouri, Columbia, Missouri 65211, USA}
\author{Matthew~Snyder}
\affiliation {Department of Physics and Astronomy, University of Missouri, Columbia, Missouri 65211, USA}
\author{Kyle~Y.~Chen}
\affiliation {Rock Bridge High School, Columbia, Missouri 65203, USA}
\author{Xiaoqian Zhang}
\affiliation {Department of Physics and Astronomy, University of Missouri, Columbia, Missouri 65211, USA}
\author{Jacob~Cook}
\affiliation {Department of Physics and Astronomy, University of Missouri, Columbia, Missouri 65211, USA}
\author{Duy~Tung~Nguyen}
\affiliation {Department of Physics and Astronomy, University of Missouri, Columbia, Missouri 65211, USA}
\author{P.~V.~Sreenivasa~Reddy}
\affiliation {Department of Physics, National Cheng Kung University, Tainan 701, Taiwan}
\author{Tay-Rong~Chang}
\affiliation {Department of Physics, National Cheng Kung University, Tainan 701, Taiwan}
\author{Pawel~J. Kowalczyk}
\affiliation {Department of Solid State Physics, Faculty of Physics and Applied Informatics, University of Lodz,
90-236 Lodz, Pomorska 149/153, Poland}
\author{Simon~A. Brown}
\affiliation {The MacDiarmid Institute for Advanced Materials and Nanotechnology, School of Physical and Chemical Sciences,
University of Canterbury, Private Bag 4800, Christchurch 8140, New Zealand}
\author{Tai-Chang~Chiang}
\affiliation {Department of Physics, University of Illinois at Urbana-Champaign, 1110 West Green Street, Urbana, Illinois 61801-3080, USA}
\affiliation {Frederick Seitz Materials Research Laboratory, University of Illinois at Urbana-Champaign, 104 South Goodwin Avenue, Urbana, Illinois 61801-2902, USA}
\author{Shengyuan A. Yang}
\affiliation {Research Laboratory for Quantum Materials, Singapore University of Technology and Design, Singapore 487372, Singapore}
\author{Guang~Bian}\email{biang@missouri.edu}
\affiliation {Department of Physics and Astronomy, University of Missouri, Columbia, Missouri 65211, USA}

\newpage

\begin{abstract}

The discovery of graphene has stimulated enormous interest in two-dimensional (2D) electron gas with linear band structure. 2D Dirac materials possess many intriguing physical properties such as high carrier mobility and zero-energy Landau level thanks to the relativistic dispersion and chiral spin/pseudospin texture. 2D Dirac states discovered so far are exclusively pinned at high-symmetry points of the Brillouin zone, for example, surface Dirac states at $\overline{\Gamma}$ in topological insulators Bi$_2$Se(Te)$_3$ and Dirac cones at $K$ and $K'$ in graphene. In this work, we report the realization of 2D Dirac states at generic $k$-points in antimony atomic layers with phosphorene structure ($i.e.$ $\alpha$-antimonene). The unpinned nature enables versatile ways to control the locations of the Dirac points in momentum space. In addition, dispersions around the unpinned Dirac points exhibit intrinsically anisotropic behaviors due to the reduced symmetry of generic momentum points. These properties make the $\alpha$-antimonene films a promising platform for exploring interesting physics in unpinned 2D Dirac fermions that are distinct from the conventional Dirac states in graphene.

\end{abstract}

\maketitle

Dirac states with linear band dispersion and vanishing effective mass have been discovered in condensed matter materials such as graphene, topological insulators, and bulk Dirac/Weyl semimetals \cite{Novoselov2005, Kim2005, Kane2010, RevModPhys.90.015001}. The topological phase of Dirac fermion states leads to many exotic physical properties such as half-quantized quantum Hall effects \cite{Castro2009}. Two-dimensional (2D) Dirac materials such as graphene have been considered as a cornerstone for the development of next-generation electronic devices. Only a few materials have been proved experimentally to host 2D Dirac states, including graphene \cite{Novoselov2005, Kim2005} and surfaces of topological insulators \cite{Kane2010, RevModPhys.83.1057}. The Dirac states in the 2D systems are mostly pinned at high-symmetry points of the Brillouin zone, such as $K$($K'$) of graphene and $\overline{\Gamma}$ of Bi$_2$Se$_3$. Recently, it has been proposed that multiple 2D Dirac states emerge at generic momentum points in the low-energy spectrum of group-Va few-layers with phosphorene-like lattice structure\cite{Lu2016}. In this case, highly anisotropic cone is a character of the unpinned Dirac point, because the generic $k$ point has much reduced symmetry compared to high-symmetry points. The unpinned nature makes the Dirac nodes movable in momentum space, e.g., by lattice strains. An energy gap can be opened by bringing together a pair of Dirac nodes. Spin-orbit coupling (SOC) of the system can also induce gaps at the Dirac nodes. All these properties lead to tunable transport properties of Dirac states, which are unavailable in conventional 2D Dirac materials. Therefore, the Dirac states in puckered group-Va few-layers are of a type distinct from those in graphene and offer new insights into the Dirac fermion physics at low dimensions. 

In this work, we report the observation of multiple unpinned Dirac states near the Fermi level in single layer (1L) and double layer (2L) antimony (Sb) films in the phosphorene structural phase. A group-Va pnictogen atom typically forms three covalent bonds with its neighbors. In the 2D limit, two allotropic structural phases, the orthorhombic phosphorene-like phase \cite{Li2014, Liu2014} and the hexagonal honeycomb-like phase \cite{Ji2016, Bian2017}, are allowed by this requirement. The two phases are referred to as $\alpha$ and $\beta$-phases, respectively, in the literature.  In the family of group-Va elements, P, As, Sb and Bi can form phosphorene-like structures\cite{Li2014, Liu2014, Liu2015, PhysRevB.91.085423, antimony2015, APL2006, PhysRevB.80.245407, Kowalczyk2013, Kundu2021}. In this work, we focus on the phosphorene-like $\alpha$-antimonene ($\alpha$-Sb for short). As shown in Figs.~1(a, b), in the lattice of $\alpha$-Sb, the Sb atoms have a strong $sp^3$-hybridization character and thus the three Sb-Sb bonds form a tetrahedral configurations. This results in two atomic sublayers with a vertical separation comparable to the bond length. In each atomic plane, the bonding between Sb atoms forms zig-zag chains along the $y$-direction. The unit cell, marked by the blue rectangle in Fig.~1(b), has a four-atom basis (two in each atomic plane). The space group of the lattice is $D_{2h}(7)$, which includes an inversion center $i$, a vertical mirror plane $\sigma_{v}$ perpendicular to $\hat{y}$,  two two-fold rotational axes $c_{2y}$ and $c_{2z}$, and a glide mirror that is parallel to the $x$-$y$ plane and lies in the middle between the two atomic planes. The glide mirror reflection is composed of a mirror reflection and an in-plane translation by (0.5$a$, 0.5$b$) \cite{Kowalczyk2020}. Previous scanning tunneling microscope (STM) and transport measurements on $\alpha$-Sb suggest the existence of linear electron bands with a fairly small effective mass near the Fermi level \cite{Shi2019}. However, a detailed spectroscopic characterization of the low-energy electronic band structure of $\alpha$-Sb is still lacking. In this work, we examine the electronic band dispersion close to the Fermi level by the method of angle-resolved photoemission spectroscopy (ARPES) and identify the unpinned Dirac states that are located at generic momentum points.

In our experiment, $\alpha$-Sb was grown on SnSe substrates in an ultrahigh vacuum environment. The SnSe crystals were cleaved $in~situ$ and provided an atomically flat surface for the growth of $\alpha$-Sb. The crystallographic structure is shown in Figs.~1(a, b). The in-plane lattice constants are $a =$ 4.65~\AA~and $b =$ 4.35~\AA~in the $x$ and $y$ directions, respectively, while the in-plane nearest-neighbor bond length is 2.90~\AA.  The two atomic sublayers are perfectly flat according to the first-principles lattice relaxations. For a single layer (1L) of $\alpha$-Sb, the vertical spacing between the two atomic sublayers is 2.79~\AA. For a double-layer (2L) $\alpha$-Sb film, the vertical distance between the two atomic sublayers within each $\alpha$-Sb layer is 2.89~\AA~while the spacing between the two layers of $\alpha$-Sb is 3.18~\AA. Figs.~1(c, d) show the STM image of two $\alpha$-Sb/SnSe samples. The first sample consists of mainly 1L $\alpha$-Sb islands, as shown in Fig.~1(c). A line-mode reconstruction (Moir{\'e} pattern) can be seen on the $\alpha$-Sb surface, as marked by the red dashed lines. The height profile taken along the red arrow (shown in Fig.~1(e)) indicates that the height of the 1L $\alpha$-Sb island on the SnSe surface is 6.1~\AA. The second sample possesses both 1L and 2L domains as shown in Fig.~1(d). The atom-resolved images taken from the 1L and 2L domains (Figs.~1(f) and 1(g)) clearly demonstrate the rectangular surface unit cell of $\alpha$-Sb. The height profile taken along the green arrow in Fig.~1(d) shows the height of the second layer is 6.8~\AA, which is slightly larger than that of the first layer due to the lattice relaxation.  
 
To study the electronic band structure, we performed first-principles calculations for the band structure of 1L and 2L $\alpha$-Sb films. The ABINIT package \cite{Gonze2009, Gonze2005} and a plane-wave basis set were utilized in the calculations. The energy cut was 400 eV. Relativistic pseudopotential functions constructed by Hartwigsen, Goedecker, and Hutter (HGH) were used \cite{PhysRevB.58.3641}. The calculated band structure of 1L $\alpha$-Sb is shown in Fig.~2. In the absence of SOC, the bottom conduction band and the top valence band at $\bar{\Gamma}$ are separated by an energy gap of ~0.3 eV, as shown in Fig.~2(a) The two bands are dominated of the $p_z$-orbital character near the zone center\cite{Lu2016}. Between $\bar{\Gamma}$ and $\bar{X}_1$, there is a small hole pocket at the Fermi level, which is generated by the overlap of a pair of bands of mainly the $p_{x,y}$-orbital character\cite{Lu2016}. To see this pocket more clearly, we plot the zoom-in band structure along the lines of `cut1' and `cut2' marked in Fig.~2(c). The conduction and valence bands cross each other and leave a pair of nodal points, which lie 0.15 eV above the Fermi level, as shown in Figs. 2(d, e). It is worth noting that  the Dirac nodes are located at generic momentum points, as schematically shown in Fig.~2(c).  In the absence of SOC, the Dirac nodes are stable due to the protection by the spacetime inversion symmetry $iT$ (where $i$ is space inversion and $T$ is time reversal symmetry), which enforces a quantized Berry phase $\theta_B$ for each Dirac point. The Berry phase along a closed loop $\ell$ encircling each Dirac node is defined as follows,
\begin{equation}
\theta_{B} = \oint_{\ell} \bm{A}_{\bm{k}}\cdot {\rm{d}}\bm{k} = \pm \pi,
\end{equation}
where $\bm{A}_{\bm{k}}$ is the berry connection of the occupied valence bands \cite{Lu2016}. On the other hand, in the presence of SOC, the number of bands is doubled due to the spin degrees of freedom.  Gap-opening terms are allowed in the Hamiltonian to lift the band degeneracy at Dirac points. This can be seen in Figs.~2(b) and 2(f, g). Though the hole pockets remain largely unchanged at the Fermi level, the conduction and valence bands are separated by an SOC-induced energy gap of ~0.07 eV. Below the Fermi level, the lower part of the gapped Dirac cone remains nearly linear. 

The band dispersion close to the Dirac nodes in absence of SOC  can be described by the linear Hamiltonian,
\begin{equation}
\widetilde{H}({\bm k})={v_x}{k_x}{\sigma_y}+\omega{v_y}({k_y}-\omega {k_0}){\sigma_x},
\end{equation}
where $k_x$ is measured from the location of the Dirac points, $\omega=\pm 1$ indicates the opposite chirality of the two Dirac nodes along `cut2' in Fig.~2(c), $k_0$ measures the separation between the two Dirac nodes in $k_y$ direction, $\sigma_{x,y}$ are the Pauli matrices for pseudospin, and $v_{x,y}$ is the group velocity at the Dirac point in $k_x$ and $k_y$ directions, respectively.  According to the calculated band structure in Fig.~2, $v_x$ = 8.35 $\times 10^5$ m/s and $v_y$ = 4.08 $\times 10^5$ m/s, indicating a highly anisotropic Dirac cone. To further examine the unpinned nature of the Dirac bands, we calculated the band structure under uniaxial lattice strains. The lattice constant $b$ in $y$ direction is changed by $-$4\%, $-$1\% and +2\%, where the `+' and `$-$' signs correspond to lattice expansion and compression, respectively. The results are plotted in Figs.~2(h-n). Under the lattice stain of $-$4\%, the conduction and valence bands are separated in energy, and thus no Dirac nodes are formed, as shown in Figs.~2(h, i). At the critical lattice strain of $-$1\%, the conduction and valence bands touch each other, leading to a quadratic Dirac cone, in which the band dispersion is linear in $k_x$ direction and quadratic in $k_y$ direction. The quadratic Dirac cone signals a topological transition between a semimetallic phase and a band insulator \cite{PhysRevB.80.153412}. The quadratic  band dispersion occurs when two Dirac points merge in a two-dimensional crystal. This can be seen from the fact that increasing the lattice constant $b$ from the critical value makes the quadratic Dirac node split into two linear Dirac nodes, as demonstrated in Figs.~2(e, m). The separation between the two Dirac nodes is sensitive to the magnitude of lattice strains. The arrows in Fig.~2(n) indicates the movement of Dirac nodes when the lattice is further expanded in $y$ direction. These results reveals two prominent features of unpinned Dirac states: (1) the Dirac nodes can freely move in the momentum space under perturbation without breaking the lattice symmetry of the system; and (2) the dispersions of Dirac bands are intrinsically highly anisotropic due to the reduced symmetry at the location of Dirac points compared to high-symmetry points. 
 
The ARPES result taken from the 1L $\alpha$-Sb sample is shown in Fig.~3. The photon energy is 21.2~eV. There are three prominent features on the Fermi surface, namely, one electron pocket at $\bar{\mathrm{\Gamma}}$ and two hole pockets near $\bar{\mathrm{X}}_{1}$  (see Fig.~3(a)). The band spectrum taken along the $\bar{\mathrm{X}}_1$-$\bar{\Gamma}$-$\bar{\mathrm{X}}_1$ direction is plotted in Fig~3(c). The calculated band dispersion is overlaid on the ARPES spectrum for comparison. The theoretical bands agree with the ARPES spectrum, especially in showing the linear bands between  $\bar{\Gamma}$ and $\bar{\mathrm{X}}_1$. We note that the molecular beam epitaxy (MBE) sample is slightly electron-doped due to the charge transfer from the SnSe substrate to the $\alpha$-Sb overlayer, and thus the Fermi level of the calculated bands was shifted to match the ARPES spectrum. In the iso-energy contour taken at 0.1~eV below the Fermi level (Fig.~3(b)), there are only a pair of circular pockets sitting close to $\bar{\mathrm{X}}_1$, which are the cross-sections of the lower Dirac cone. The zoom-in spectra along `cut1' and `cut2' are plotted in Fig.~3(d), showing clearly the $quasi$-linear band dispersion from the lower part of the gapped Dirac cone. 

We also calculated the band structure of 2L $\alpha$-Sb. The results are summarized in Fig.~4. In the absence of SOC (Fig.~4 (a)), there are multiple band crossings along the $\bar{\Gamma}$-$\bar{\mathrm{X}}_1$ direction near the Fermi level (which can also be seen in the zoom-in band along `cut3' in Fig.~4(c)). This is because the number of $p_{x,y}$-orbital dominated bands is doubled in the 2L film. In addition, we see another band crossing along the $\bar{\Gamma}$-$\bar{\mathrm{X}}_2$ direction. Unlike the nodal points close to the $\bar{\Gamma}$-$\bar{\mathrm{X}}_1$ line, the bands at this new nodal point are of mainly the $p_z$ orbital character. In Figs.~4(b, d),  the zoom-in band along `cut1' and `cut2' (marked in Fig.~4(m)) show that this new Dirac point is located at a generic momentum point between $\bar{\Gamma}$ and $\bar{\mathrm{X}}_1$ and sit at 0.2~eV below the Fermi level. The Dirac band is gapless in the absence of SOC\cite{Lu2016}. We also plot the bands along `cut4' and `cut5' in Figs.~4(e, f) to show the Dirac states close to the line of $\bar{\Gamma}$-$\bar{\mathrm{X}}_1$. Figure~4(e) shows a highly anisotropic Dirac cone. The Dirac node is in the $\bar{\Gamma}$-$\bar{\mathrm{X}}_1$ line and at 0.2 eV above the Fermi level. The two branches of the Dirac cone are nearly degenerate in the direction of `cut4'. On the other hand, the bands along `cut5' (Fig.~4(f)) show a pair of Dirac nodes at 0.4~eV above the Fermi level, which are generated by the small overlap between the conduction and valence bands. The dispersion is similar to the Dirac bands of 1L $\alpha$-Sb in the absence of SOC. The locations of these unpinned Dirac nodes are schematically summarized in Fig.~4(m). The band dispersions calculated with SOC are plotted in Figs.~4(g-l). Energy gaps are opened at all Dirac nodes marked in Fig.~4(m). Compared to the case of 1L $\alpha$-Sb, the size of the energy gap in 2L $\alpha$-Sb is much smaller. For example, the energy gap in the Dirac bands along `cut1' is 6~meV as shown in the zoom-in plot in Fig.~4(h). Therefore, the whole band structure resembles the gapless Dirac cone in the absence of SOC.   

We measured the band structure of the 2L $\alpha$-Sb sample shown in Fig.~1(e). We note that the sample consists of 1L and 2L domains. The Fermi surface contour is plotted in Fig.~5(a). The nodal features at generic momentum points can be easily identified. Besides those nodal features, there is a bright pocket located at the zone center, which is actually from the 1L domains of the samples. The band spectrum along the $\bar{\mathrm{X}}_2$-$\bar{\Gamma}$-$\bar{\mathrm{X}}_2$ direction (Fig.~5(b)) exhibits the linear Dirac band dispersion, which is in good agreement with the calculated bands. This can be more clearly seen in the zoom-in spectra along `cut1' and `cut2' plotted in Fig.~5(c). The Dirac node is at 0.2~eV below the Fermi level. The small energy gap opened by SOC (6~meV according to the band calculation) is beyond the energy resolution of our ARPES measurements (15 meV). Figure~5(d) shows the nodal features and linear bands along `cut3-5'. The experimental results are consistent with the calculated band dispersion (the red lines),  demonstrating the existence of the unpinned Dirac states in 2L $\alpha$-Sb.

In summary, our ARPES measurements and first-principles calculations unambiguously demonstrated that $\alpha$-Sb hosts unpinned Dirac states in both 1L and 2L cases. The Dirac nodes are protected by the spacetime inversion symmetry in the absence of SOC. The 2D Dirac nodes at generic $k$-points are unpinned and have highly anisotropic dispersions, which are experimentally confirmed in this study for the first time. The unpinned nature enables versatile ways, such as lattice strains, to control the locations and the dispersion around the Dirac points. SOC of the system induces energy gaps at the Dirac nodes. The gap size is 70~meV for the Dirac node of 1L $\alpha$-Sb between $\bar{\Gamma}$ and $\bar{\mathrm{X}}_1$ and 6~meV for the Dirac node of 2L $\alpha$-Sb between $\bar{\Gamma}$ and $\bar{\mathrm{X}}_2$.  The observed $quasi$-linear band dispersion resembles those of gapless 2D Dirac states. For example, the circular iso-energy contours shown in Fig.~3(b) have the same winding number as the gapless Dirac cones. Dirac points with a small gap also have interesting effects, which are useful for transport and optical applications \cite{Zhou2008, Xu2012, PhysRevB.81.195203}. In the family of group-Va elements, P, As, Sb and Bi can form phosphorene-like structures\cite{Li2014, Liu2014, Liu2015, PhysRevB.91.085423, antimony2015, APL2006, PhysRevB.80.245407, Kowalczyk2013, Kundu2021}. The conduction and valence bands of P and As monolayers are far apart in energy (1.2 eV for P and 0.4 eV for As). Therefore, no Dirac states could form near the Fermi level. By contrast, the conduction and valence bands of Sb and Bi layers overlap in energy and can generate band crossings near the Fermi level. The size of the SOC-induced energy gaps depends on the effective strength of SOC of the system. The strength of atomic spin-orbit coupling is proportional to the fourth power of the atomic number ($\propto Z^4$). As a result, Bi has a much larger SOC coupling than Sb, resulting in sizable SOC-induced energy gaps in the bands of monolayer Bi\cite{Bian2017, Lu2016}. Therefore, $\alpha$-Sb atomic layers demonstrated in this work provides the best opportunity, compared with other members of group-Va films, for exploring novel properties of unpinned 2D Dirac fermions.

\bibliographystyle{apsrev4-1}
\bibliography{Sb_unpinned}

\newpage

\begin{figure}
\includegraphics[width=1.0\linewidth]{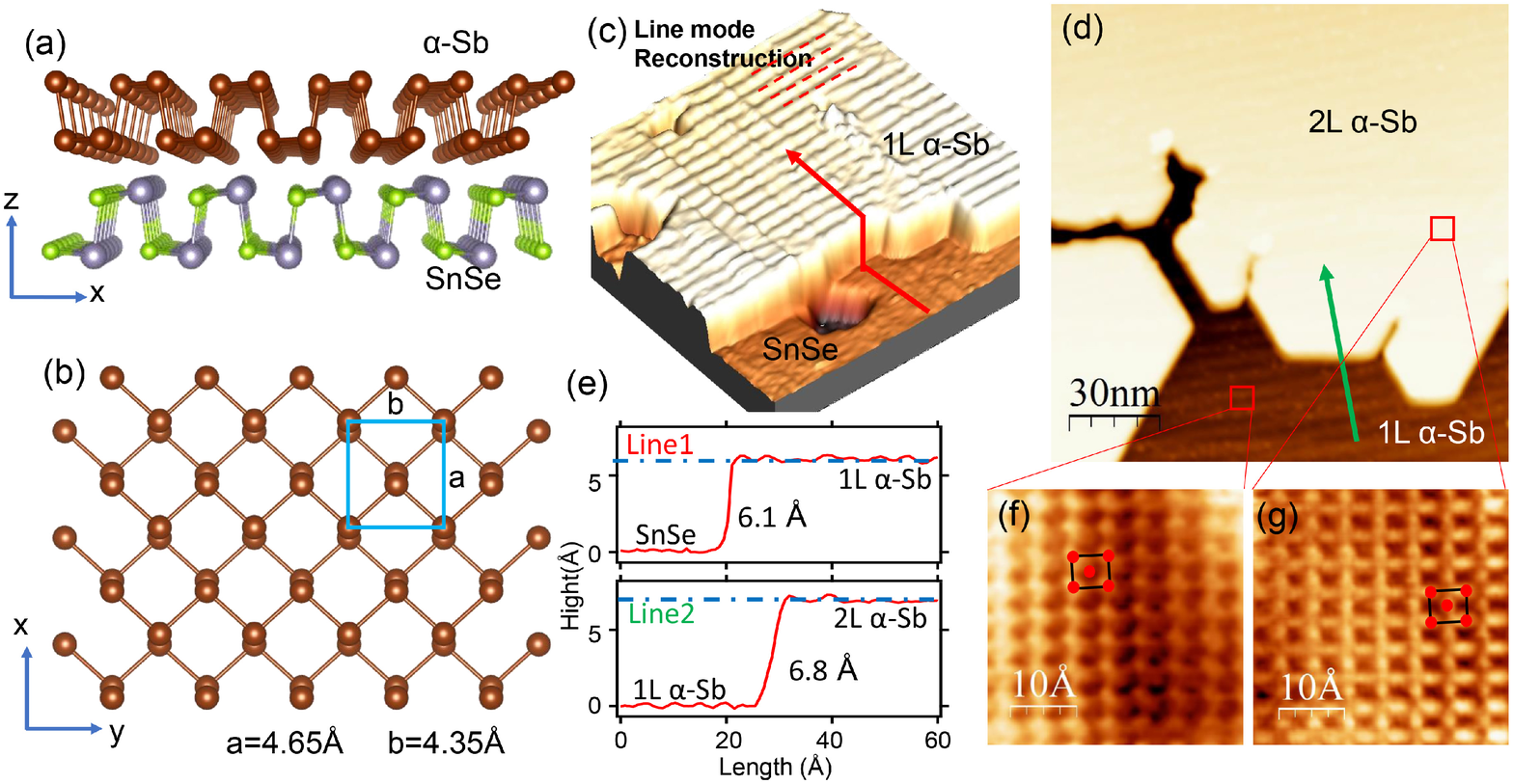}
\caption{(a) Side view of $\alpha$-Sb/SnSe lattice structure. (b) Top view of $\alpha$-Sb lattice structure. The unit cell is indicated by the blue rectangular box. (c) STM image of $\alpha$-Sb grown on SnSe substrate. (d) STM image showing both 1L and 2L domains. (e) The height profiles taken along the red arrow in (c) and the green arrow in (d). (f, g) Atom-resolved STM images taken from 1L and 2L domains. 
}%
\end{figure}

\newpage

\begin{figure}
\includegraphics[width=1.0\linewidth]{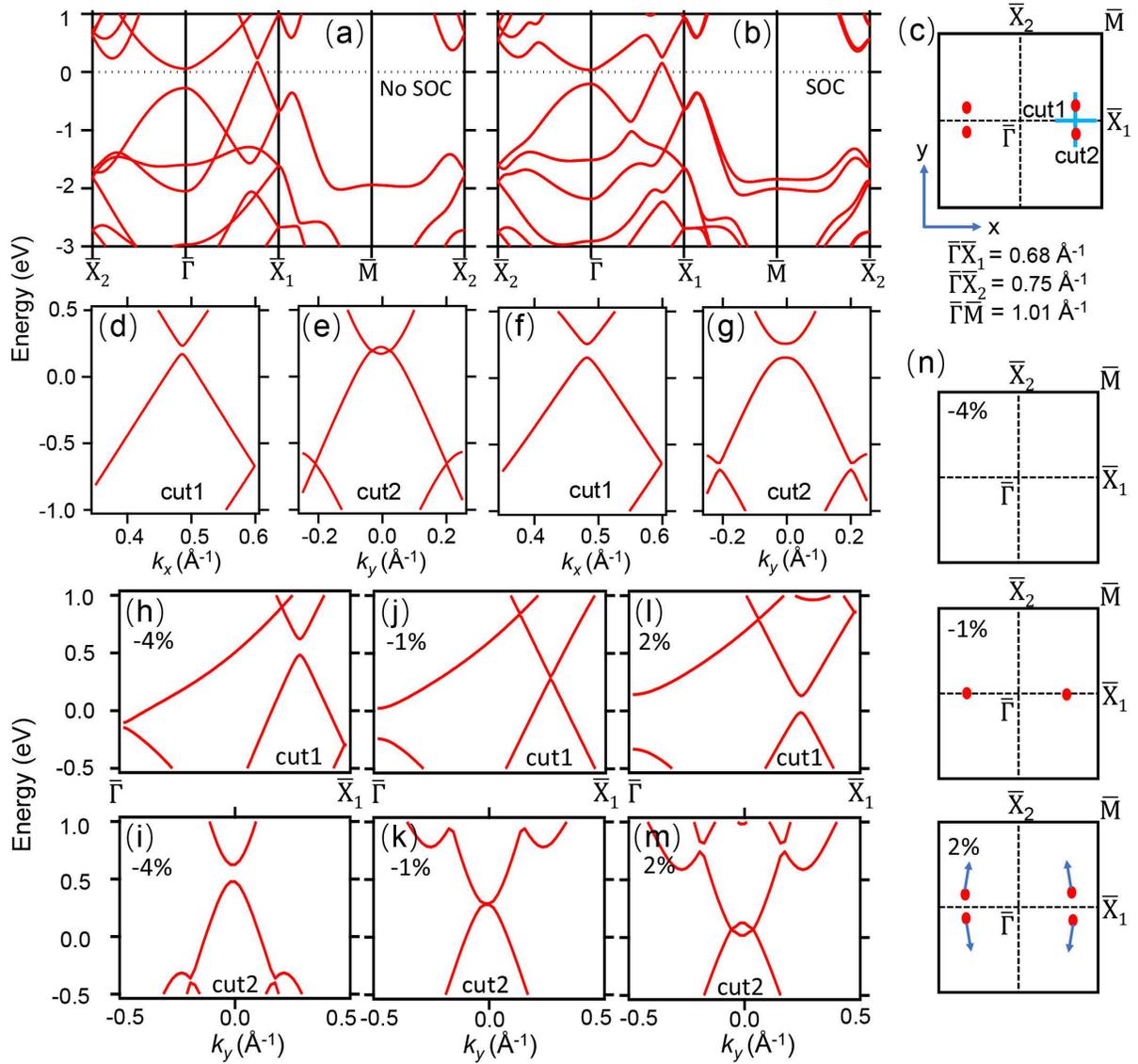}
\caption{(a, b) The band structure of 1L $\alpha$-Sb calculated without and with the inclusion of SOC. (c) The Brillouin zone of 1L $\alpha$-Sb. (d, e) The calculated band dispersion along `cut1' and `cut2' (marked in (c)) without the inclusion of SOC. (f, g) Same as (d, e), but calculated with the inclusion of SOC. (h-n) Band structure (without SOC) of 1L $\alpha$-Sb under uniaxial lattice strains along $y$ direction. (h, i) Band dispersion along `cut1' and `cut2' (marked in (c)) with $-$4\% lattice strain in $y$ direction. (j, k) and (l, m) Same as (h, i), but for lattice strains in $y$ direction of $-$1\% and +2\%, respectively. (n) The locations of Dirac points in the Brillouin zone. The blue arrows indicate the movement of Dirac points as the lattice constant $b$ increases.}%
\end{figure}

\newpage
\begin{figure}
\includegraphics[width=1.0\linewidth]{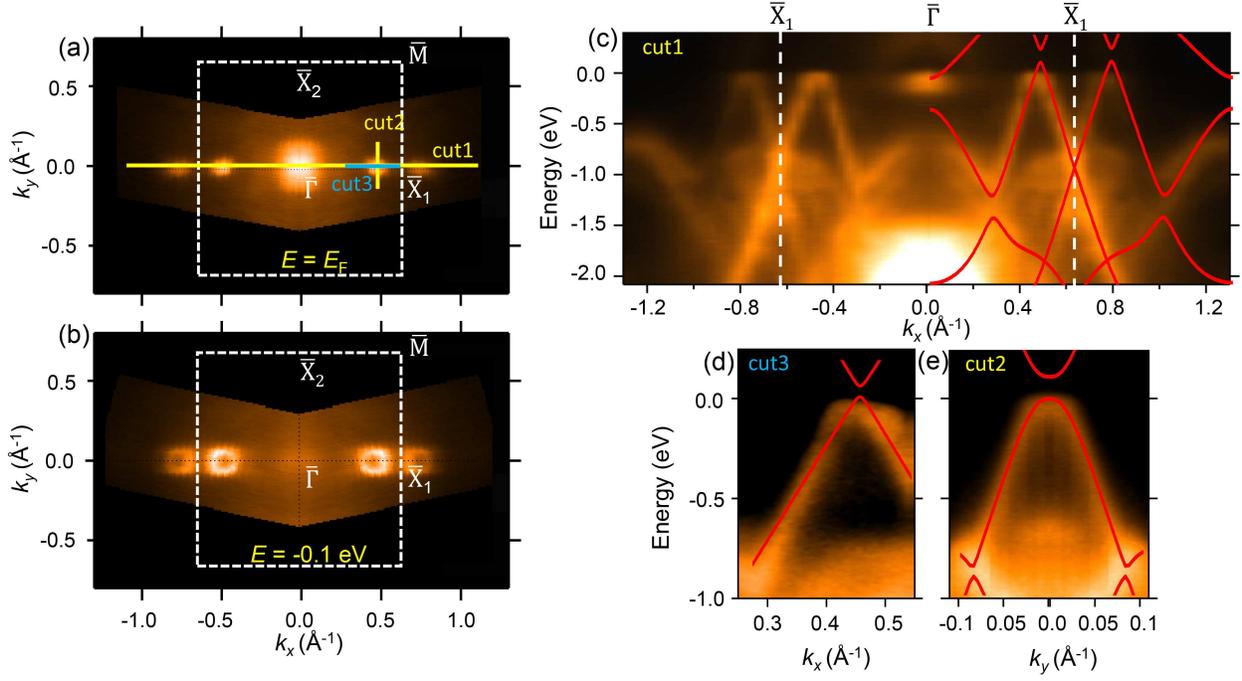}
\caption{ (a) Fermi surface taken from the 1L $\alpha$-Sb sample by ARPES. (b) ARPES iso-energy contour taken at $E=-0.1$ eV. (c)  ARPES spectrum taken along $\bar{\mathrm{X}}_1$-$\bar{\mathrm{\Gamma}}$-$\bar{\mathrm{X}}_1$. (d-e) Zoom-in spectra taken along `cut3' and `cut2'  (marked in (a)), respectively.}%
\end{figure}

\newpage
\begin{figure}
\includegraphics[width=1.0\linewidth]{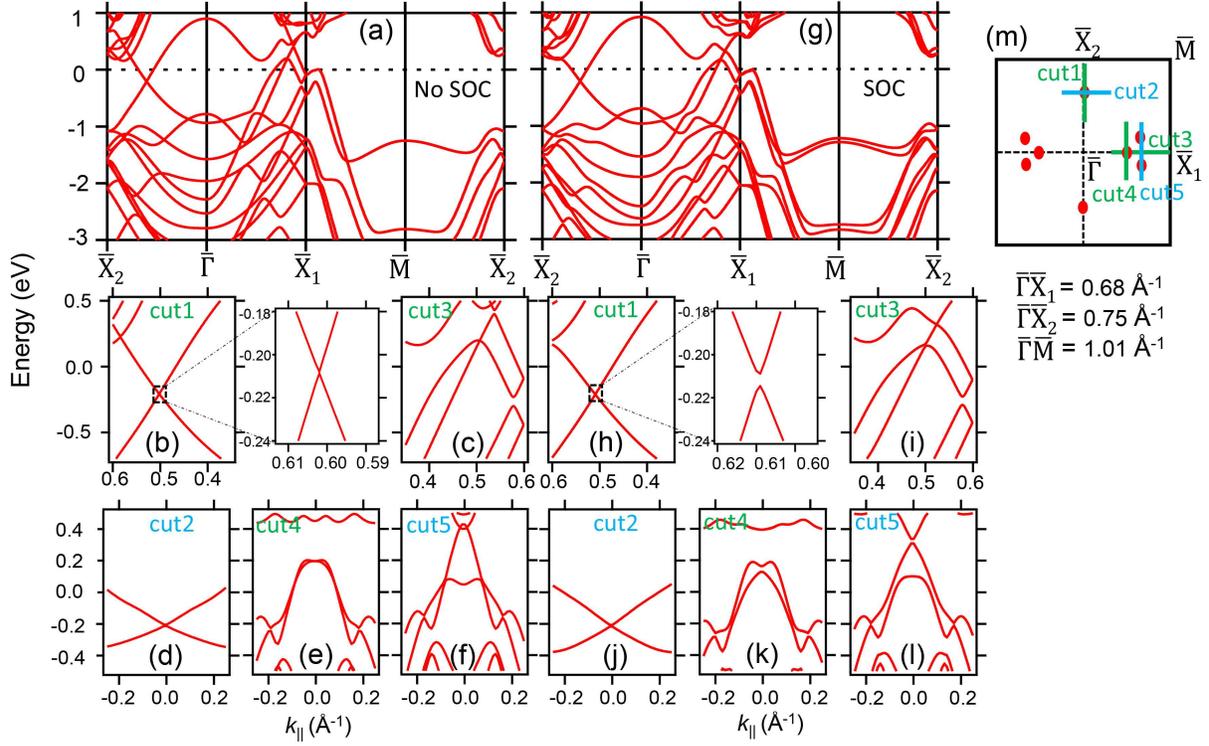}
\caption{(a) The band structure and density of states of 2L $\alpha$-Sb calculated without the inclusion of SOC. (b-f) The calculated band dispersions along `cut1-5', respectively. (g-l) Same as (a-f), but calculated with the inclusion of SOC. (m) The Brillouin zone of 2L $\alpha$-Sb. The Dirac points in the case without SOC are marked by red dots.}%
\end{figure}

\newpage
\begin{figure}
\includegraphics[width=1.0\linewidth]{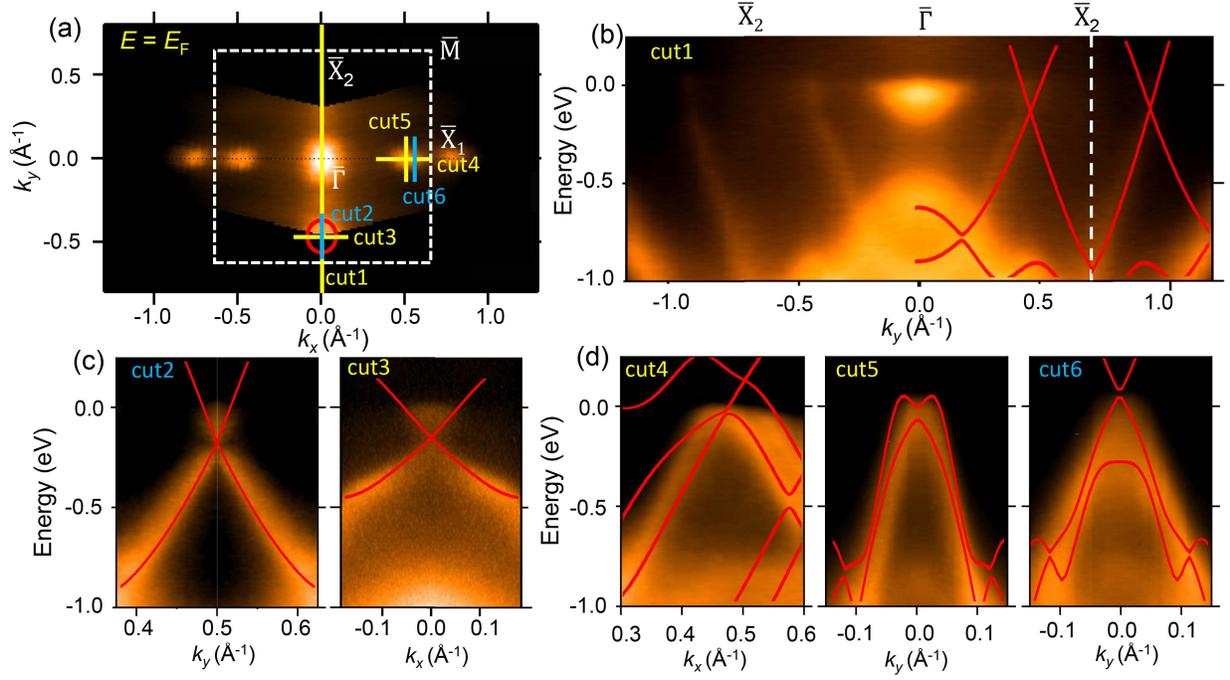}
\caption{(a) Fermi surface taken from the 2L $\alpha$-Sb sample by ARPES. (b) ARPES spectrum taken along $\bar{\mathrm{X}}_2$-$\bar{\mathrm{\Gamma}}$-$\bar{\mathrm{X}}_2$. (c,d) ARPES spectra taken along `cut2-6' (marked in (a)), respectively. The spectrum along `cut2' is symmetrized to visualize the Dirac band structure.}%
\end{figure}

\end{document}